\documentclass[aps,pra,twocolumn,longbibliography,superscriptaddress]{revtex4-1}
\usepackage{graphicx}
\usepackage{dcolumn}
\usepackage{bm}
\usepackage{hyperref}
\usepackage{amsmath}
\usepackage{braket} 
\usepackage{MnSymbol}
\usepackage{color}

\newcommand{\dens}{$\rho_0\lambda^{3}$}
\newcommand{\Gsup}{$\Gamma_\mathrm{sup}$}
\newcommand{\Gsub}{$\Gamma_\mathrm{sub}$}
\newcommand{\tsup}{$\tau_\mathrm{sup}$}
\newcommand{\tsub}{$\tau_\mathrm{sub}$}

\newcommand {\INLN} {Universit\'e C\^ote d'Azur, CNRS, INPHYNI, France}
\begin{document}


\title{Decay dynamics in the coupled-dipole model}

\author{M. O. Ara\'ujo}
\affiliation{\INLN}
\affiliation{CAPES Foundation, Ministry of Education of Brazil, Bras\'ilia, DF 70040-020, Brazil}
\author{William Guerin}
\email{william.guerin@inphyni.cnrs.fr}
\affiliation{\INLN}
\author{Robin Kaiser}
\affiliation{\INLN}

\begin{abstract} 
Cooperative scattering in cold atoms has gained renewed interest, in particular in the context of single-photon superradiance, with the recent experimental observation of super- and subradiance in dilute atomic clouds. Numerical simulations to support experimental signatures of cooperative scattering are often limited by the number of dipoles which can be treated, well below the number of atoms in the experiments. In this paper, we provide systematic numerical studies aimed at matching the regime of dilute atomic clouds. We use a scalar coupled-dipole model in the low excitation limit and an exclusion volume to avoid density-related effects. Scaling laws for super- and subradiance are obtained and the limits of numerical studies are pointed out. We also illustrate the cooperative nature of light scattering by considering an incident laser field, where half of the beam has a $\pi$ phase shift. The enhanced subradiance obtained under such condition provides an additional signature of the role of coherence in the detected signal.
\end{abstract}

\maketitle

\section{Introduction}

Since the seminal work by Dicke in 1954~\cite{Dicke:1954}, a vast range of phenomena has been studied in the context of light emission and scattering by an ensemble of $N$ two-level systems~\cite{Friedberg:1973,Feld:1980,Gross:1982,Lagendijk:1996,Labeyrie:2008,Manassah:2012,Bienaime:2013,Guerin:2017a,Kupriyanov:2017}. More recently, the properties of such situations when only one excitation at most is present in the system has been studied, both theoretically and experimentally (``single-photon superradiance''~\cite{Scully:2006,Scully:2009,Frowis:2017}). Some of the theoretical work is based on the study of an effective Hamiltonian, investigating either the escape rates of photons from the system~\cite{Akkermans:2008}, related to the imaginary part of the effective Hamiltonian, or the eigenvalues of the complete effective Hamiltonian~\cite{Rusek:1996,Rusek:2000,Pinheiro:2004}, with in particular the prediction of a localization transition in the scalar model, absent in a more complete vectorial model~\cite{Skipetrov:2014,Bellando:2014,Skipetrov:2016c}.

Experimental studies of cooperative effects in atom-light interaction however typically involves an incident laser beam, either driving the system to a steady state, or realizing a pulsed excitation to study the dynamics of collective effects. The experimental signatures studied so far include the momentum transfer to the center of mass of the atomic cloud~\cite{Courteille:2010,Bienaime:2010,Chabe:2014,Bachelard:2016} or the light scattered either in the backward direction~\cite{Labeyrie:1999}, the forward direction~\cite{Kwong:2015,Jennewein:2016,Roof:2016}, or at different angles~\cite{Pellegrino:2014,Guerin:2016a,Araujo:2016}.

Theoretical investigations of interference effects in multiple scattering around the backward direction (coherent backscattering) can be performed using an approximate diagrammatic approach~\cite{Jonckheere:2000}. For forward scattering, steady-state properties as well as the dynamics after the switch off of the driving field can be well understood by using a description based on the average refractive index of the cloud~\cite{Kwong:2015,Roof:2015,Roof:2016}, as long as the atomic sample remains at low density~\cite{Jennewein:2016,Jenkins:2016b}.
However, these efficient theoretical approaches do not allow describing scattering at a random angle, as used in~\cite{Pellegrino:2014,Guerin:2016a,Araujo:2016}. Nevertheless, in the single scattering limit, an alternative approach has been proposed that allows for analytical predictions of the scattering off axis~\cite{Kuraptsev:2017}.

One more universal approach to light scattering in any direction, in both single and multiple scattering limits, is the so-called coupled-dipole model~\cite{Javanainen:1999,Svidzinsky:2010}. This approach, which can be derived in the low-excitation limit either in a quantum framework or from classical scattering, requires a numerical solution of $N$ coupled equations. The near-field dipole-dipole coupling as well as the polarization of the electromagnetic radiation can be taken care of, but a simplified model consists in a scalar description of the dipole-dipole interaction. Even though not exact, this scalar approximation has the merit of having allowed identifying the important role of near-field coupling in the problem of Anderson localization~\cite{Skipetrov:2014,Bellando:2014}, not discussed in the context of single parameter scaling~\cite{Abrahams:1979}. Another advantage of the scalar model is that it allows the simulation of large optical depth without density-related effects. Indeed, as one typically can use up to $N=10^4$ atoms in the numerical simulations and the on-resonance optical depth scales as  $b_0\propto N/(k_0R)^2$, the simulation of $b_0\approx 20$ corresponds to a size of the cloud of $k_0R\approx 20$, where $k_0=2\pi/\lambda$ is the wavenumber corresponding to the atomic transition and $R$ the size of the atomic cloud. For a fixed number of atoms, larger values of $b_0$ are thus simulated by smaller values of $k_0R$. This comes along with larger spatial densities $\rho$ and smaller interatomic distances $d \sim 1/\rho^{1/3}$, yielding for the above choice of parameters values of $k_0d\sim 1$. For such densities, the near-field term of the vectorial coupled-dipole model significantly affects the results. The simulation of the dilute limit realized in the experiments is thus difficult. A good compromise is thus to use the scalar model, keeping in mind that this model features a phase transition when the density reaches values of $\rho \lambda^3\approx 20$~\cite{Skipetrov:2014,Bellando:2014}. However, even below this critical density, where one expect the long-range coupling to dominate over the local coupling, we have noticed that for a randomly-filled cloud of particles, occasionally two atoms will be located at very short distance and pair physics can become visible~\cite{Rusek:2000,Skipetrov:2011,Bellando:2014}. As such pair physics is less probable in the dilute atomic cloud, we have added an exclusion volume around each atom to avoid close distances and related effects. A precise simulation of scaling laws as expected in the large atom number limit is therefore delicate and the scope of this paper is to provide the systematic studies we have performed to explore super- and subradiance in dilute clouds of cold atoms. We note that the experiments~\cite{Guerin:2016a,Araujo:2016} are performed with rubidium atoms, having a degenerate ground state with multiple Zeeman sublevels. The additional complexity related to this degeneracy is largely out of reach for present theories, even though some attempts to take care of effects related to this more complex internal structure of the atoms have been made recently~\cite{Jenkins:2016b,Hebenstreit:2017}.

As a further study of cooperativity, we illustrate the role of coherence in subradiance by applying a recently proposed idea~\cite{Scully:2015} to a realistic experimental geometry. Using a properly phased excitation, we show that the time-dependent scattering is indeed sensitive to the phase profile of the laser field and, more importantly for experimental aspects, that such a phased excitation can induce an increase of the subradiant fraction of light by almost one order of magnitude. Controlling the addressed states of the full Hilbert space by engineered steering is a splendid illustration of the potential of cooperative scattering in such open (quantum) systems.

This paper is organized as follows: in section \ref{sec_model}, we review the scalar coupled-dipole model we use to describe the interaction between an atomic cloud and a laser beam.
In section \ref{sec_results}, we discuss in more detail the impact of the exclusion condition to avoid close atomic pairs and we present scaling laws of super- and subradiance obtained in this framework. In section \ref{sec_phased_cloud}, we study a phased excitation of the atomic cloud, yielding in particular an increase of subradiant scattering. Finally, we summarize our findings in section \ref{sec_conclusion}.

\section{The coupled-dipole model}
\label{sec_model}

\subsection{Coupled-dipole equations}

\begin{figure}[t]
\centering
\includegraphics[scale=1]{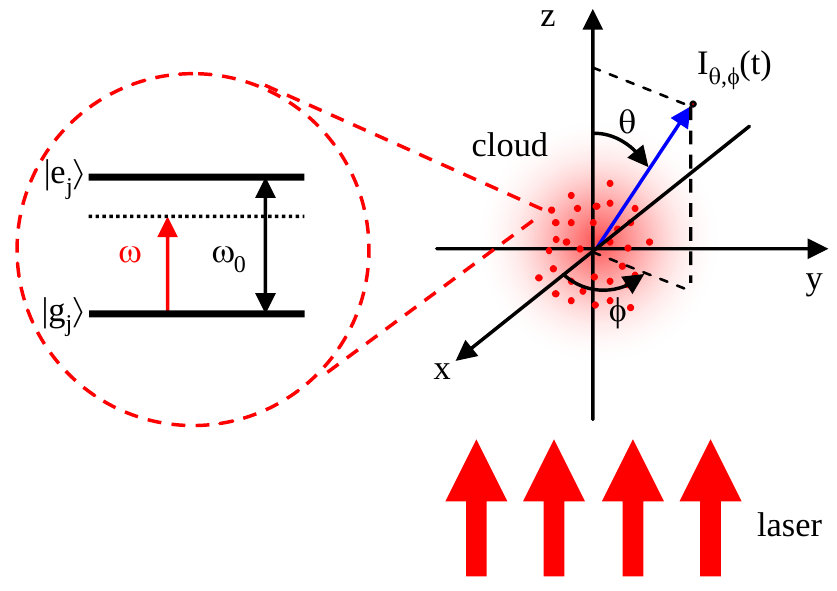} 
\caption{The physical system. A spherical Gaussian cloud of identical atoms interacts with a monochromatic plane wave with frequency $\omega$ and vacuum modes. The wave vector $\bm{k}_0$ of the plane wave is set along the $z$ direction. Each atom is a two-level system with resonance frequency $\omega_0$. The angles $\theta$ and $\phi$ defines the spherical coordinates.}
\label{fig_cloud}
\end{figure}

Even though it is possible to derive the equations describing the evolution of coupled dipoles in the low excitation limit from classical equations, a quantum formalism is often conveniently used.
We thus consider a set of $N$ identical atoms interacting with a laser beam and vacuum modes (Fig. \ref{fig_cloud}). Each atom $j$ is a non-degenerate two-level system, with $\Ket{g_j}$ (resp. $\Ket{e_j}$) labeling the ground (resp. excited) state. The resonance frequency of the atomic transition is $\omega_0$, the natural lifetime of the excited state is $\tau_0$ and the natural decay rate is $\Gamma=1/\tau_0$. The atoms are considered motionless and distributed at random positions $\bm{r}_j$ in a given distribution that models the geometry of the sample. The laser beam is described by a classical monochromatic plane wave of amplitude $E_0$, wave vector $\bm{k}_{0}$ and frequency $\omega$.

Following~\cite{Courteille:2010,Bienaime:2011} we write the coupled-dipole equations in the scalar approximation as
\begin{equation}
\dot{\beta}_{j}=
\left( i\Delta - \dfrac{\Gamma}{2} \right) \beta_{j} -
\dfrac{i\Omega}{2} e^{i\bm{k}_{0}\cdot\bm{r}_{j}}-
\dfrac{\Gamma}{2} \sum\limits_{j'\neq j}
\dfrac{e^{ik_{0}r_{jj'}}}{ik_{0}r_{jj'}}
\beta_{j'}\, ,
\label{betas}
\end{equation}
where $r_{jj'}=\left| \bm{r}_j - \bm{r}_{j'} \right| $ is the relative distance between the atoms $j$ and $j'$, $\Omega = dE_{0}/\hbar $ is the Rabi frequency associated to the laser drive, and $\Delta=\omega-\omega_{0}$ is the laser detuning from the atomic resonance.
Here $\beta_j(t)$ are the time-dependent amplitudes of dipoles $j$.
Eq. (\ref{betas}) is valid for weak driving fields, i.e., the solution in the linear-optics regime. This amount to approximating the quantum solution to the first order in $\Omega$, which is valid for
$s(\Delta)\ll 1 $, where  $s(\Delta)=2\Omega^2/(\Gamma^2+4\Delta^2)$ is the saturation parameter. We note that testing this assumption quantitatively, in particular in regard to the long-lived subradiant modes, would require a full quantum treatment~\cite{Pucci:2017}.
In a quantum framework, $\beta_j(t)$ can be considered as the optical coherence of the atom $j$ and the wave function of the atom ensemble reads
\begin{equation}
\Ket{\Psi(t)}=\alpha(t)\Ket{G}+\sum\limits_{j=1}^N \beta_j(t)\Ket{j} \, ,
\label{Psi}
\end{equation}
where $\Ket{G}\equiv\Ket{g_1...g_N}$ is the ground state for all atoms and $\Ket{j}\equiv\Ket{g_1...e_j...g_N}$ are the $N$ single-atom excited states for each atom $j$.

\subsection{Light scattered by the atomic cloud}
\label{sec_emited_power}

The electric field at a position $\textbf{r}$ far from the atomic cloud is related to the atomic dipoles $ \beta_j(t)$ by
\begin{equation}
E(\textbf{r}, t) \propto
\dfrac{e^{-i\omega_{0}(t-r/c)}}{r}
\sum\limits_{j=1}^{N} e^{-ik_0\hat{r}\cdot \bm{r}_{j}} \beta_j(t) \, ,
\end{equation}
where $\hat{r}=\bm{r}/r$ defines the direction of the detection from the center of the atomic cloud. We thus obtain the angular emitted intensity $I \propto |E|^2$ by
\begin{equation}
I(\hat{r},t) \propto \dfrac{1}{r^{2}}
\left| \sum\limits_{j=1}^{N} e^{-ik_0\hat{r}\cdot\bm{r}_{j}} \beta_{j}(t) \right|^{2} \, .
\label{I_rt}
\end{equation}

In spherical coordinates, the emitted intensity in the direction
$(\theta$,$\phi)$ (see Fig.~\ref{fig_cloud})
can thus be written as
\begin{equation}
I_{\theta,\phi}(t) \propto \left| \sum\limits_{j=1}^{N} e^{-ik_0 f_j(\theta,\phi)} \beta_{j}(t) \right|^{2} \, ,
\label{P_theta_phi_t}
\end{equation}
where $f_j(\theta,\phi)=x_j\sin\theta\cos\phi+y_j\sin\theta\sin\phi+z_j\cos\theta$ and $\bm{r}_j = (x_j, y_j, z_j)$ are the coordinates of the atom $j$.
We thus first solve the coupled-dipole equations (\ref{betas}) and from the solution of $\beta_j(t)$ we compute the angular- and time-resolved scattered intensity using Eq.~(\ref{P_theta_phi_t}).

With a plane wave laser propagating along the $z$ axis with $\bm{k}_0=k_0\hat{z}$ and for an atomic cloud with revolution symmetry around the $z$ axis, we can integrate the angular intensity distribution along $\phi$ to obtain
\begin{equation}
I_{\theta}(t) \propto
\int_{0}^{2\pi}
I_{\theta,\phi}(t) d\phi
=
\int_{0}^{2\pi}
\left| \sum\limits_{j=1}^{N} e^{-ik_0 f_j(\theta,\phi)} \beta_{j}(t) \right|^{2}
d\phi \, .
\label{P_theta_t}
\end{equation}
This integration along $\phi$ yields lower fluctuations for finite numerical resolution. However, this procedure does not allow to study speckle-like fluctuations of the scattered light. Also, as we will see in section \ref{sec_phased_cloud}, for a laser drive with a phase profile which does not respect the revolution symmetry around the $z$ axis, one cannot use the integration over $\phi$ to study super- and subradiance by a `phased' excitation.


The total emitted power $ P(t) $ is the power emitted by the atoms integrated over all directions:  $P(t)=\int I(\bm{r},t) d\bm{r} $ with $d\bm{r} = r^{2}\sin\theta dr d\theta d\phi$. By substituting Eq.~(\ref{I_rt}), we obtain
\begin{equation}
P(t) \propto
\int_{0}^{\pi}\int_{0}^{2\pi}
\left| \sum\limits_{j=1}^{N} e^{-ik_0 f_j(\theta,\phi)} \beta_{j}(t) \right|^{2}
\sin\theta d\theta d\phi.
\label{P_integral}
\end{equation}

In this paper, we are mainly interested in the decay dynamics after the extinction of the driving laser and thus consider only times $t\geq0$, where $t=0$ corresponds to the switch off of the driving field. For $t\geq0$, $\Omega=0$, one can show that Eq.~(\ref{P_integral}) yields (see Appendix)
\begin{equation}
P(t) \propto -\dfrac{d}{dt} \sum\limits_{j=1}^N \left| \beta_{j}(t) \right|^{2} \, ,
\label{P_derivative}
\end{equation}
which can be interpreted as the energy transfer from the dipoles to the light.
This simplified expression for the time-dependant total scattered light allows for faster numerics, convenient for some initial explorations, but as we show below, it is unable to provide the full scaling laws for superradiance, where angle-dependent effects can be prominent.

\section{Decay dynamics for a spherical Gaussian cloud}
\label{sec_results}

The atomic system is modeled as a spherical cloud with a Gaussian probability distribution $\rho(\bm{r}_j) = \rho_0 e^{-|\bm{r}_j|^2/2R^2}$, where $R$ is the rms size of the cloud. The resonant optical thickness is $b_0=2N/(k_0 R)^{2}$ and the peak density \dens\ $=(2\pi)^{3/2}N/(k_0R)^{3}$. Note that here we use the definition of the optical depth for a scalar model, different from what should be used in a vectorial model [$b_0^\mathrm{(v)}=3N/(k_0R)^{2}$], where the polarization of the light yields an on-resonant scattering cross section of $3\lambda^2/2\pi$. The comparison to the optical depth in the experiment is furthermore different, as for atoms in a statistical mixture of the ground states, the on-resonant optical depth is reduced by a degeneracy factor $g=\frac{2F'+1}{3(2F+1)}$, which, for the $F=2\rightarrow F'=3$ transition of rubidium 87 used in~\cite{Guerin:2016a,Araujo:2016} gives a reduction of the on-resonant optical depth of $g\approx 0.47$.

Typical experimental values \cite{Guerin:2016a,Araujo:2016} are $\lambda\sim 1\mu$m, $N\sim10^{9}$ and $R\sim 1$~mm, yielding  $b_0\sim10-100$ and \dens\ $\sim 10^{-2}$. With \dens$\ll 1$, the cloud is very dilute, allowing to consider near-field effects as negligible. However, it is very hard to simulate a cloud with $10^{9}$ atoms on available computers. So in practice we use $N\sim10^{3}$--$10^{4}$, but we require $b_0\sim 10$, in order to have similar optical depths in the simulations as in the experiments. As a consequence of smaller $N$ in the simulation, the simulated density increases to \dens\ $\sim 1$--10.
Moreover, the computed intensities should be averaged for many different configurations of the positions of the atoms, in order to remove residual fluctuations, important for the small number of atoms used in simulations. In this article we have averaged over 100 realizations.

\subsection{Impact of an exclusion volume} \label{sec.exclusion}

Our procedure for simulations consists in generating a spherical Gaussian cloud by choosing randomly the positions of the $N$ atoms from a Gaussian distribution of rms size $R$. However, as close atomic pairs may occur with such a random choice of positions, with a probability increasing with the atomic density, and considering that such pairs \cite{DeVoe:1996,Rusek:2000,Skipetrov:2011,Bellando:2014} lead to strong two-body super- and subradiant decay absent in very dilute clouds, we choose to add an exclusion volume around each atom. After drawing the atomic positions in the cloud, we look for pairs of atoms with distance $k_0r_{jj'} \leq 3$ and change the corresponding positions until the condition $k_0r_{jj'}>3$ if fulfilled for all pairs. This numerical value for the exclusion volume corresponds to a distance where the two-body super- and subradiant decay rates, $\Gamma_2 = \Gamma \left[1 \pm \sin(k_0 r_{jj'})/k_0r_{jj'} \right]$, become close to single-atom physics.

\begin{figure}[t]
\centering
\includegraphics[scale=1]{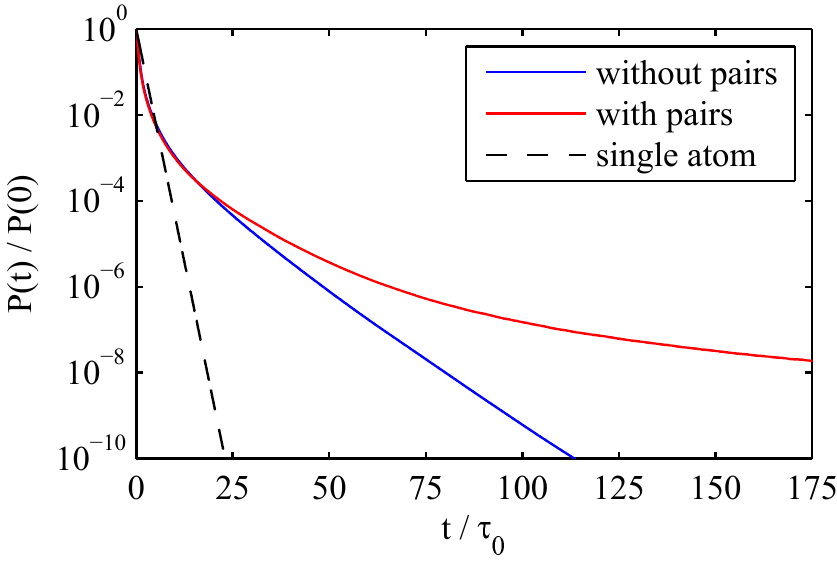}
\caption{Comparison between the total emitted fluorescence for a spherical cloud with and without pairs (i.e., without and with an exclusion volume, respectively) for $k_0R\approx 12.9$, $b_0\approx 7.55$ and \dens\ $\approx 4.6$. The exclusion-volume condition is $k_0r_{jj'}<3$ for the distance between all pairs of atoms $j$ and $j'$.
The detuning is $\Delta=10\Gamma$ and the curves are obtained after averaging over $100$ different configurations of the atomic positions.}
\label{fig_pairs}
\end{figure}

In order to illustrate the pair effect in the decay dynamics, we plot in Fig.~\ref{fig_pairs} the total scattered light (Eq.~\ref{P_derivative}) for a cloud with and without using an exclusion volume. The cloud without pairs was generated from initial target parameters $b_0=10$ and \dens$=7$ for the Gaussian distribution (with corresponding values of $N=633$ and $k_0R=11.3$). After the exclusion condition is fulfilled, almost $50\%$ of the $N$ atoms had their position changed, however their distribution remains Gaussian with a good approximation, but with a larger size: $k_0R$ increased to $12.9$, implying that $b_0$ and \dens\ have to be recalculated. That gives $b_0=7.55$ and \dens$=4.6$.
Note that this increase of size comes along with introducing some correlation in the positions, which might induce some spurious density effects.
To compare this situation to a cloud without exclusion volume, we choose those final values for $b_0$ and \dens, only setting randomly the positions of the atoms without the exclusion condition.
As shown in Fig. (\ref{fig_pairs}), the presence of atomic pairs increases the subradiant amplitude at long times. This can be understood by the very long lifetimes of the two-atom subradiant state for very small distances.

\subsection{Scaling of the super- and subradiance with $b_0$}

\begin{figure}[b]
\centering
\includegraphics[scale=1]{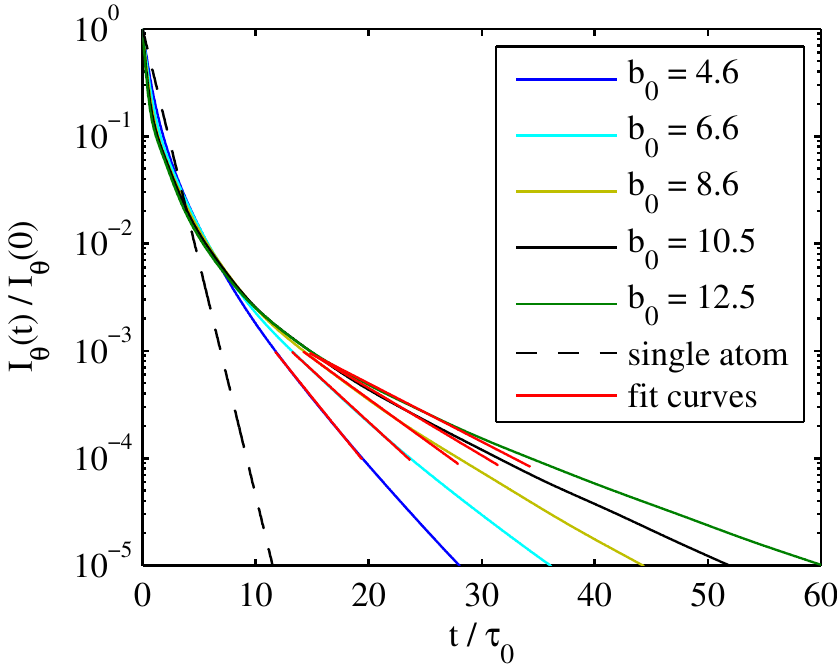}
\caption{
Decay of the scattered light in a direction $\theta=45^{\circ}$ from the laser direction, computed from Eq.~(\ref{P_theta_t}), for atomic samples of several $b_0$ and constant density \dens$=4.6$. The laser detuning is $ \Delta=10\Gamma$. The red curves are exponential fits in the interval $I_\theta(t)/I_\theta(0)\in[10^{-4},10^{-3}]$ that allow extracting the subradiant decay rate \Gsub. A similar fit (not showed) is done in the interval $t\in[0,0.2]\tau_0$ to extract the superradiant decay rate \Gsup. The black dashed line is the decay curve for a single atom.}
\label{fig_fits}
\end{figure}

From decay curves similar as the one shown in Fig.~\ref{fig_pairs}, one can extract the super and subradiant decay rates \Gsup, \Gsub (or their associated time constant \tsup $=$ \Gsup $^{-1}$ , \tsub $=$ \Gsub $^{-1}$) by using an exponential fit in the appropriate range. For superradiance, the fit interval must be set at short time after the laser switch off, in order to get the first time constant. Here, we have used the fitting interval $t\in[0,0.2]\tau_0$. For subradiance, the fit interval must be set long after the switch off of the laser, but, as discussed in the following (Sect.~\ref{sec.long_time}), the precise choice is somewhat arbitrary. Here we have chosen to fit the subradiant decay rate in the interval given by the relative amplitude of the signal, $I_\theta(t)/I_\theta(0) \in [10^{-4}, 10^{-3}]$, similar to what has been used in the experiment~\cite{Guerin:2016a}. Examples of decay curves and associated fits are shown in Fig.~\ref{fig_fits}. Faster (superradiant) and slower (subradiant) decays compared to the single-atom decay are well visible. We have chosen the emission angle $\theta=45^\circ$ for this illustration, but one can check that the decay is similar for other directions, except in the exact forward direction $\theta=0$, where a strong superradiant forward scattering lobe is present~\cite{Scully:2006,Bromley:2016,Roof:2016}, and consequently the relative amplitude of the subradiant decay is much lower.

The duration of the pulse of the driving laser before switch off can also affect the results, as studied in detail in~\cite{Kuraptsev:2017}. Here we consider only long-pulse excitation so that the steady-state is reached before the switch off. We have numerically used a duration $T_\mathrm{pulse} = 100 \tau_0$ and have checked that starting from the steady-state solution before switching off the laser does not modify the results.


The behavior of the super- and subradiant decay rates as a function of the parameters of the cloud (atom number, size, optical depth or density) has been the subject of extensive discussions. In the case of a low-density sample driven at large detuning, the collective steady state is essentially the timed-Dicke state~\cite{Scully:2006,Bienaime:2013,Guerin:2017b}. The corresponding superradiant decay rate \Gsup\ has been analytically computed for various geometries~\cite{Mazets:2007,Svidzinsky:2008,Svidzinsky:2008b,Courteille:2010,Friedberg:2010,Prasad:2010} and it has been shown that it is proportional to the resonant optical depth $b_0$. Based on numerical investigations, it has been argued that the same apply for subradiance, with a characteristic time \tsub\ evolving linearly with $b_0$~\cite{Bienaime:2012}. This result has been checked numerically and experimentally in~\cite{Guerin:2016a,footnoteJMO2018}. It has been actually showed that the full decay curves depends only on $b_0$ in this regime~\cite{Guerin:2017a}.

\begin{figure}[t]
\centering
\includegraphics[scale=1]{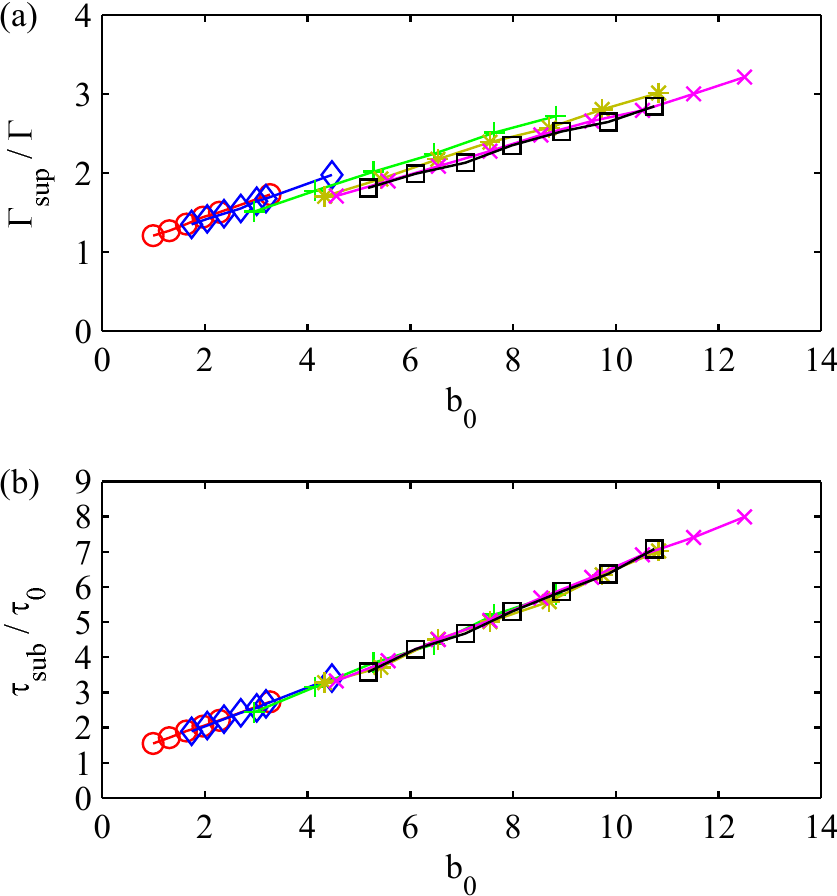}
\caption{Cooperative decay rates \Gsup\ (a) and \tsub$=$\Gsub$^{-1}$ (b) as a function of the resonant optical depth $b_0$ for different densities \dens$=0.5$ (red $\circ$), $0.9$ (blue $\diamond$), $2.5$ (green $+$), $3.7$ (yellow $\ast$), $4.6$ (magenta $\times$) and $5.3$ (black $\square$). The decay rates are obtained by fitting decay curves similar as those in Fig.~\ref{fig_fits}, computed with a laser excitation at a large detuning $\Delta = 10 \Gamma$ and a long pulse in order to reach the steady state.}
\label{fig_decay}
\end{figure}

All previous calculations and simulations, however, apply to the total decay rate of the collective state, which is mainly dominated by the forward scattering lobe. In Fig.~\ref{fig_decay} we present the superradiant decay rate \Gsup\ and subradiant decay time \tsub\ computed from the scattered light at $\theta=45^\circ$ as a function of $b_0$ for several densities, following an excitation to the steady state with a large detuning $\Delta=10\Gamma$. The chosen angle is close to the experimental configuration of~\cite{Guerin:2016a,Araujo:2016}. The computation is done for several densities, in order to check whether the density plays any role. The results of the simulations show that the density plays only a marginal role, inducing a small shift of the superradiant decay rate (Fig.~\ref{fig_decay}a) and not affecting the subradiant one (Fig.~\ref{fig_decay}b). This residual density effect might come from the correlations introduced by the exclusion volume condition and should thus be absent at the lower densities of the experiments. The main trend is clearly the linear evolution of the superradiant decay rate and subradiant time constant, following
\begin{eqnarray}
\Gamma_\mathrm{sup} & = \left( 1+ \alpha b_0 \right) \Gamma \, , \label{eq.slope_super}\\
\tau_\mathrm{sub} & = \left( 1+\beta b_0 \right) \tau_0 \, .
\end{eqnarray}
Linear fitting the low-density data gives $\alpha \approx 0.21$ and $\beta \approx 0.53$. For comparison, the only available analytical results is for the total decay rate of the time-Dicke state, which gives $\alpha = 1/8$ for a Gaussian cloud~\cite{Bienaime:2011}. We indeed recover this scaling for forward scattering, but superradiance is ``stronger'' for light scattered off-axis, as already discussed in~\cite{Araujo:2016}. Note that we do not expect any quantitative agreement with the experiments because of the complex multilevel structure of the atoms used in the experiments.

\subsection{Angular dependance of super- and subradiance}

As already mentioned, the decay dynamics in general depends on the scattering angle (Eq.~\ref{P_theta_t}). The most prominent feature of the angular dependence is the forward scattering lobe~\cite{Scully:2006,Bromley:2016}. It is visible in steady state but, as it is mainly superradiant, it also affects the temporal dynamics, with, in particular, a lower relative weight of subradiant decay in the forward direction and a slightly different slope $\alpha$ (Eq.~\ref{eq.slope_super}) for the superradiant decay~\cite{Araujo:2016}. An extensive study of the superradiant decay rate as a function of the scattering angle has recently been presented, showing a complicated dependency near the forward lobe~\cite{Kuraptsev:2017}. It also depends on the detuning of the exciting laser.

\begin{figure*}
\centering
\includegraphics[scale=1]{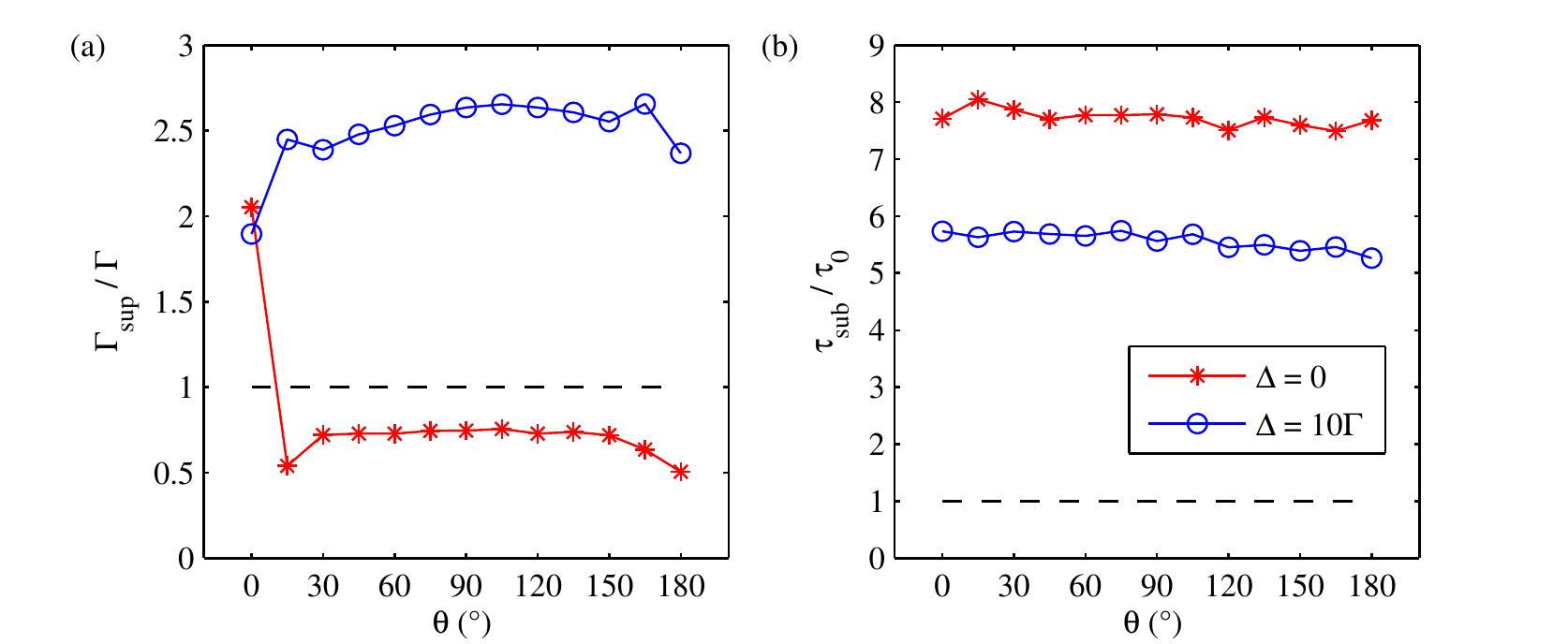}
\caption{
Cooperative decay rates \Gsup\ (a) and \tsub\ (b) as a function of the detection angle $\theta$ measured from the laser axis, where $\theta = 0$ is the forward direction and $\theta = 180^{\circ}$ is the backscattering direction. Data were calculated for on-resonance (red) and far-detuned (blue) excitation, for a sample with fixed density \dens$=4.6$ and $b_0=8.55$.
In (a), superradiance is suppressed on-resonance except in the forward direction. Off-resonance, it is faster off-axis ($\theta\neq0$) than on-axis ($\theta=0$). In (b), subradiance is isotropic for all detunings and directions. The fitting range of the subradiant decay has been set to a lower value for the point at $\theta=0$ in order to take into account the lower relative amplitude of subradiance due to the forward superradiant lobe.
}
\label{fig_angles}
\end{figure*}

In Fig.~\ref{fig_angles}(a) we summarize these findings on the superradiant decay rate as a function of the scattering angle for near-resonance and detuned excitation, and we also show the behavior of the subradiant decay in Fig.~\ref{fig_angles}(b). The resonant optical depth is fixed, $b_0 = 8.55$. At large detuning, the superradiant decay rate \Gsup\ is significantly larger off axis than exactly on axis, which is a nonintuitive feature. Except very close to the forward lobe, where \Gsup\ evolves a lot with the angle (not shown here, see~\cite{Kuraptsev:2017}), \Gsup\ hardly evolves everywhere else. On resonance, on the contrary, superradiance is visible only in the forward direction, and \Gsup$<\Gamma$ off axis, which means that superradiance is suppressed and that the light escape rate is slowed down by multiple scattering~\cite{Labeyrie:2003}. Although the collective state is different from the timed-Dicke state in this situation~\cite{Bienaime:2013}, and more generally superradiant modes are much less populated near resonance \cite{Guerin:2017b}, the forward superradiant lobe is still preserved because it comes from the diffracted light on the edge of the cloud when driven by a plane wave. For subradiance, we do not see any significant variation of \tsub\ with the emission angle, confirming initial intuition that subradiance is on average isotropic~\cite{Bienaime:2012,Guerin:2016a}. The relative amplitude of the subradiant decay is however much smaller at $\theta=0$ because of the dominant forward superradiant lobe. The slightly slower decay at resonance is due to multiple scattering, which contributes to slowing down the decay~\cite{Labeyrie:2003}. The precise interplay between subradiance and multiple scattering near resonance is the subject of studies in progress.

\subsection{Long-time limit}\label{sec.long_time}

\begin{figure}[b]
\centering
\includegraphics[scale=1]{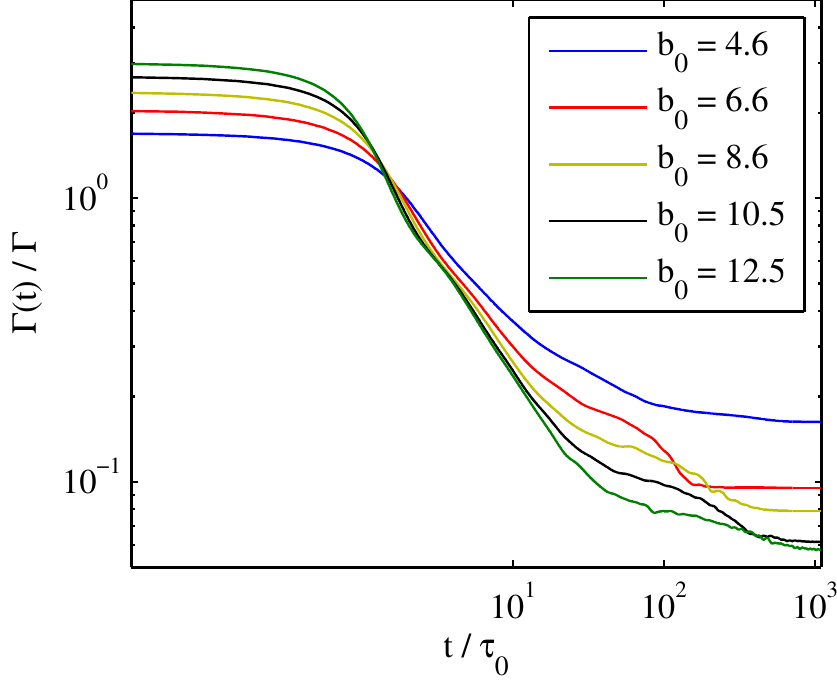}
\caption{
Temporal decay rates $\Gamma(t)$ calculated from the total fluorescence (Eq. \ref{P_derivative}), for several $b_0$, with \dens$=4.6$ and $\Delta=10\Gamma$. The decay rates still evolve even for very long times, becoming completely constant for $t\geq600\tau_0$ after the extinction of the driving laser.}
\label{fig_gamma_t}
\end{figure}

As mentioned at the beginning, the fitting interval for subradiance is somewhat arbitrary and the subradiant decay time that can be measured in an experiment is related to the noise level below which the decay is not measurable. It is thus interesting to look if the decay rate still evolves at very late times, even if those times are not accessible experimentally.
From the numerical decay curves, we can compute the instantaneous decay rate
\begin{equation}
\Gamma(t) = -\frac{d}{dt}\left[\ln P(t) \right] \, ,
\label{Gamma_t}
\end{equation}
here defined for the total scattered light (Eq.~\ref{P_integral}).

\label{sec_phased_cloud}
\begin{figure*}
\centering
\includegraphics[scale=1]{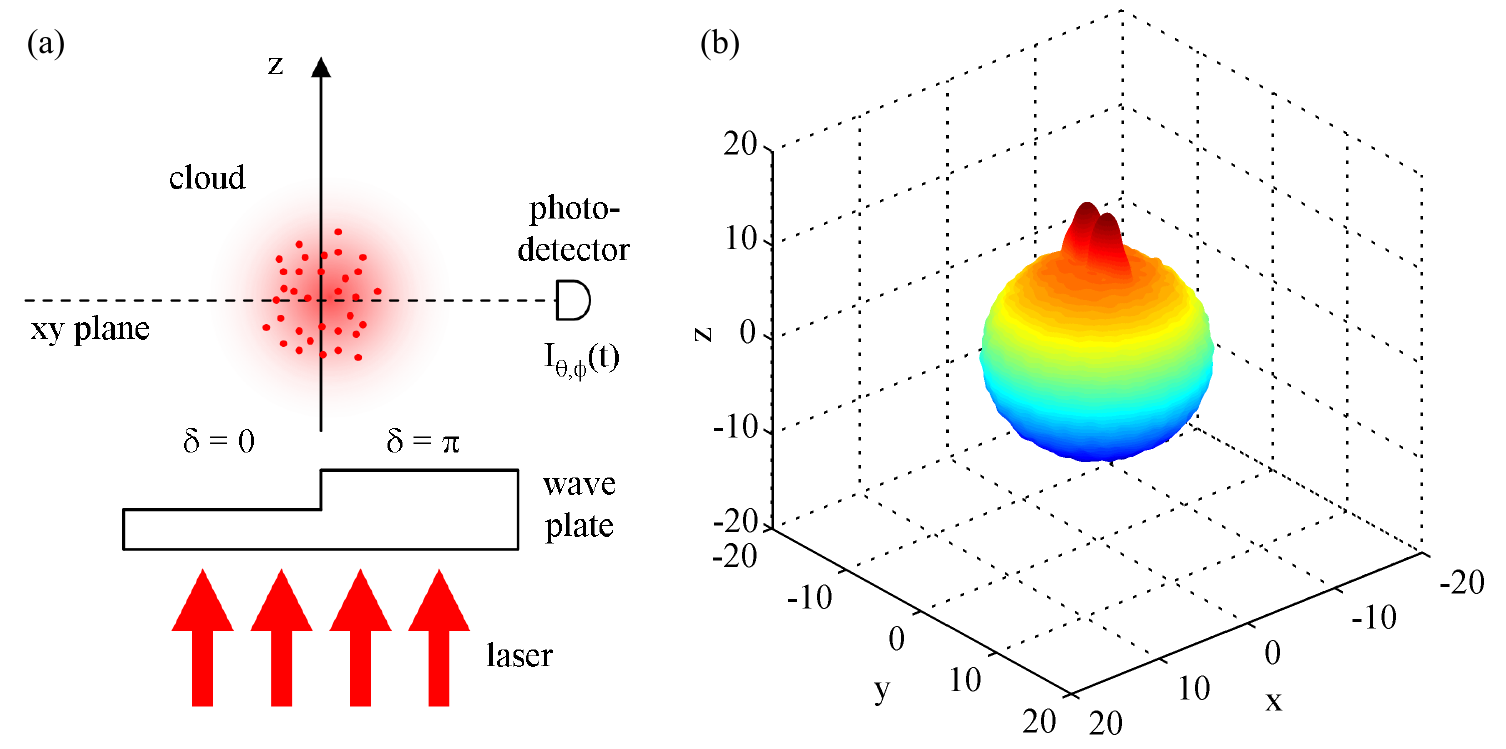}
\caption{
Phased cloud. (a) The cloud is excited by a laser propagating along the $z$ axis, which goes through a wave plate such as the half space $y>0$ of the cloud is excited with a phase shift of $\delta=\pi$ compared to the atoms in the half space $y<0$. The detector measures the intensity $I_{\theta,\phi}(t)$ computed from Eq.~(\ref{P_theta_phi_t}).
(b) Emission diagram in the steady state for the phased cloud, for $b_0=8.55$, \dens$=4.6$ and $\Delta=10\Gamma$. The single forward lobe, characteristic of superradiance with a plane wave (without phase shift), is divided into two components for the ``phased" excitation. The emission diagram is isotropic in other directions.
}
\label{fig_phased_cloud_1}
\end{figure*}

In Fig.~\ref{fig_gamma_t} we show $\Gamma(t)$ for $\Delta=10\Gamma$ and different values of $b_0$. We find that the decay rates still evolve long after the switch off, until it eventually reaches a constant value at very long time, here $t\geq600\tau_0$. We have checked that this does not depend on the duration of the exciting pulse. The final value of $\Gamma(t)$ corresponds to the smallest real part of the eigenvalue spectrum (see~\cite{Guerin:2017b}), which is very sensitive to the configuration of the positions. Moreover, the population of the corresponding mode can be negligible, which means that the signal is far below the detection threshold of any realistic experiment. The long-time limit of $\Gamma(t)$ is thus not relevant from an experimental point of view. On the other hand, this problem shows that we lack a rigorous definition of the subradiant decay rate, and we have to rely on empirical definitions to measure or numerically evaluate it.

\section{Subradiance in a phased cloud}

\begin{figure}[t]
\centering
\includegraphics[scale=1]{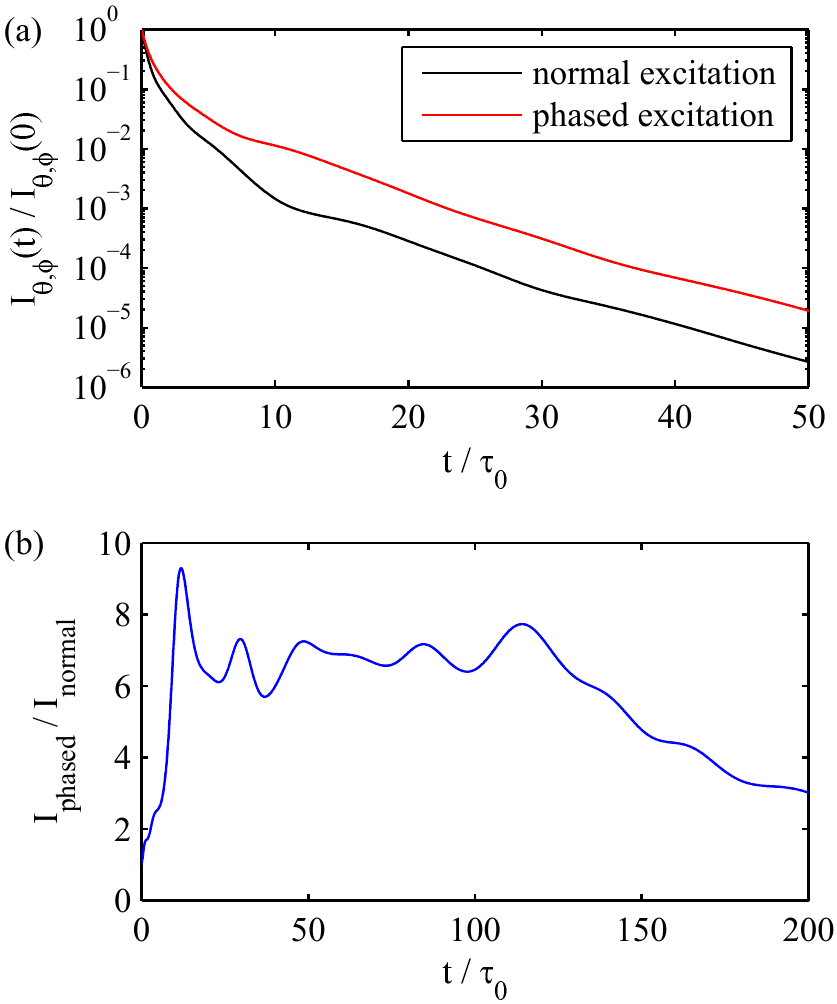}
\caption{
(a) Comparison between the normal and phased cloud for the decay dynamics. The parameters of the sample are $b_0=8.55$, \dens$=4.6$ and the laser detuning is $\Delta=10\Gamma$. Superradiance is almost completely suppressed and the relative amplitude of subradiance is increased, while the decay rate is conserved. Here the scattered intensity is calculated at $\theta=90^\circ$, $\phi=90^\circ$ (see Fig.~\ref{fig_phased_cloud_1}), but the result is similar for all $\phi$.
(b) Ratio between the scattered light from the phased and normal cloud, during the decay, as computed in (a). The phased excitation increases the relative amplitude of subradiance by a factor $\sim 6$--8.
}
\label{fig_phased_cloud_2}
\end{figure}

Subradiance is hard to detect experimentally because of its low relative level in the decay dynamics, on the order of $10^{-3}$~\cite{Guerin:2016a}, which limits the prospect of using subradiant states for quantum information processing, metrology, or transport experiments~\cite{Ostermann:2013,Scully:2015,Cai:2016,Facchinetti:2016,Asenjo:2017,Celardo:2017}. Finding a way to selectively populate the subradiant states, or at least to enhance their relative weight, is thus a relevant challenge.

Inspired by~\cite{Scully:2015}, in this section we analyze the decay dynamics of what we call a \textit{phased cloud} (Fig.~\ref{fig_phased_cloud_1}a). This is an atomic cloud where one half side is driven with a phase shift of $\delta=\pi$ compared to the atoms in the other half ($\delta=0$). This set-up can be experimentally realized by making the incoming laser to go through a wave plate of different thickness. Then, after the laser extinction, the scattered light is detected in the direction orthogonal to the phase step.

For the simulations, the driving laser is set along the $z$-axis and the phase shifter is placed such that the atoms in the half space $y>0$ are driven by the laser with the phase $k_0 z_j+\pi$ in Eq.~(\ref{betas}), while the atoms in the half space $y<0$ are driven with the phase $k_0 z_j$ as previously. As the phased-cloud system is not symmetric with respect to the $z$-axis, we use Eq.~(\ref{P_theta_phi_t}) to compute the scattered light in a given direction $\theta$, $\phi$. Fig.~\ref{fig_phased_cloud_1}b shows the emission diagram of the phased cloud in the steady state. The main effect of the phase step is to divide the forward emission into two lobes. This is due to the destructive interference between the light emitted by the atoms in the two different half spaces. A similar idea, controlling the direction of the main diffraction lobe via the phase of the driven atoms, has been discussed in~\cite{Maximo:2014}.

The remarkable property of this phased cloud is the decay dynamics for directions orthogonal to the laser axis ($\theta=90^\circ$).
Fig.~\ref{fig_phased_cloud_2}(a) compares the scattered intensity for the phased excitation and the standard excitation, i.e., without phase shifter (``normal" excitation). As previously, the data are normalized to the steady-state values before switch off, which are similar in both cases. However, after some decay time, the signal obtained with the phased excitation is significantly higher. This means that the relative weight of subradiance is enhanced by a factor $\sim 6-8$ [Fig.~\ref{fig_phased_cloud_2}(b)], which is a significant improvement for experiments. It also shows that subradiance is sensitive to the phase of the laser, which is a marked difference with radiation trapping~\cite{Labeyrie:2003}, and thus stresses that this is a coherent effect.

This improvement obtained with a very simple phase profile for the incident laser shows that there should be possible to enhance the population of the subradiant states by engineering the phase or intensity or temporal profile of the exciting laser.

\section{Summary}
\label{sec_conclusion}

In this article, we have reported a detailed numerical investigation on the decay dynamics for a dilute spherical Gaussian cloud of $N$ particles, using the coupled-dipole equations. We have discussed the effect of using an exclusion-volume condition avoiding pairs of close atoms, in order to better simulate the low-density regime in which some experiments are performed~\cite{Guerin:2016a,Araujo:2016}. It was shown that the superradiant decay rate scales linearly with the resonant optical depth $b_0$. On the contrary, the subradiant decay rate, although defined with some arbitrary considerations, is inversely proportional to $b_0$. We have also shown that while the superradiant decay rate is significatively different for the light scattered on axis and off axis, the subradiant decay rate is isotropic. Finally, the decay dynamics of a sample driven by a laser with a $0-\pi$ phase profile was studied and it was shown that this configuration increases significantly the relative weight of subradiance detected in some directions. This paves the way for the development of even more involved and efficient way to selectively populate the subradiant states, which is promising for their future use in various fields, for instance for quantum information processing, metrology, or transport experiments~\cite{Ostermann:2013,Scully:2015,Cai:2016,Facchinetti:2016,Asenjo:2017,Celardo:2017}.

\section*{Acknowledgements}


We thank Marlan Scully for insightful suggestions during the 2017 PQE Conference and Carlos M\'aximo, Romain Bachelard, Philippe Courteille, and Nicolas Piovella for useful discussions about the coupled-dipole model and simulation methods.

\section*{Funding}

We acknowledge financial support from the French Agency ANR (project LOVE, No. ANR-14-CE26-0032) and the Brazilian CAPES Foundation.



%

\appendix
\section{Proof of Equation~(\ref{P_derivative})}

The equivalence of the Eqs.~(\ref{P_integral}) and (\ref{P_derivative}) is verified by showing that both are proportional to
\begin{equation}
I = \sum\limits_{j}\sum\limits_{j'} \dfrac{\sin(k_{0}r_{jj'})}{k_{0}r_{jj'}} \beta_{j}\beta_{j'}^{*} \, .
\label{A_1}
\end{equation}
For simplicity we have dropped the $j,j'=1,...,N$ in the summation.

Let us to start with Eq.~(\ref{P_integral}). We rewrite it as
\begin{equation}
P(t) \propto
\sum\limits_{j}\sum\limits_{j'}
\int_{0}^{2\pi}\int_{0}^{\pi}
e^{-ik_0\hat{r}\cdot(\textbf{r}_{j}-\textbf{r}_{j'})} \beta_{j}\beta_{j'}^{*}
\sin\theta d\theta d\phi \, .
\end{equation}
By choosing a coordinate frame such that $ \hat{r}\cdot(\textbf{r}_{j}-\textbf{r}_{j'}) = r_{jj'}\cos\theta $~\cite{Jackson:book}, and noting that the integral over $\phi$ gives $2\pi$, we have
\begin{eqnarray}
P(t) &\propto& 2\pi \sum\limits_{j}\sum\limits_{j'} \beta_{j}\beta_{j'}^{*} \int_{0}^{\pi} e^{-ik_{0}r_{jj'}\cos\theta} \sin\theta d\theta  \nonumber \\
&=& 2\pi \sum\limits_{j}\sum\limits_{j'} \beta_{j}\beta_{j'}^{*} \int_{-1}^{1} e^{-ik_{0}r_{jj'}u} du \nonumber \\
&=& 4\pi \sum\limits_{j}\sum\limits_{j'} \dfrac{\sin(k_{0}r_{jj'})}{k_{0}r_{jj'}} \beta_{j}\beta_{j'}^{*} \, ,
\label{I_1}
\end{eqnarray}
i.e., proportional to Eq.~(\ref{A_1}). Note that Eq.~(\ref{I_1}) is valid for all $t$ and not only during the decay ($t>0$).

Now we turn to Eq.~(\ref{P_derivative}). We start by expanding the derivative in the right-hand side by using the Kronecker delta:
\begin{eqnarray}
f(t) & \equiv & -\dfrac{d}{dt} \sum\limits_{j} \left| \beta_{j}(t) \right|^{2} = -\dfrac{d}{dt} \sum\limits_{j}\sum\limits_{j'} \beta_{j}\beta_{j'}^{*}\delta_{jj'} \nonumber \\
& = & -\sum\limits_{j}\sum\limits_{j'} \delta_{jj'} \left( \dot{\beta}_{j}\beta_{j'}^{*} + \beta_{j}\dot{\beta}_{j'}^{*} \right) \, .
\label{I_2}
\end{eqnarray}
The terms $\dot{\beta_j}$ are replaced by the coupled-dipole equation (\ref{betas}) and its complex conjugate, rewritten with the term $(-\Gamma/2)\beta_{j}$ absorbed into the summation, i.e.,
\begin{equation}
\dot{\beta}_{j}=
i\Delta \beta_{j} -
\dfrac{i\Omega}{2} e^{i\textbf{k}_{0}\cdot\textbf{r}_{j}}-
\dfrac{\Gamma}{2} \sum\limits_{j'}
V_{jj'} \beta_{j'}\, .
\end{equation}

After the substitutions, we have
\begin{eqnarray}
f(t) &=& -\sum\limits_{j}\sum\limits_{j'} \delta_{jj'} \left( \left[ i\Delta \beta_{j} -
\dfrac{i\Omega}{2} e^{i\textbf{k}_{0}\cdot\textbf{r}_{j}}-
\dfrac{\Gamma}{2} \sum\limits_{m}
V_{jm} \beta_{m} \right] \beta_{j'}^{*} \right. \nonumber\\
& & \quad + \beta_{j} \left. \left[ -i\Delta \beta_{j'}^{*} +
\dfrac{i\Omega}{2} e^{-i\textbf{k}_{0}\cdot\textbf{r}_{j'}}-
\dfrac{\Gamma}{2} \sum\limits_{n}
V_{j'n}^{*} \beta_{n}^{*} \right]  \right) \nonumber \\
&=& \dfrac{\Gamma}{2} \sum\limits_{j}\sum\limits_{j'} \delta_{jj'} \left( \sum\limits_{m}
V_{jm} \beta_{m}\beta_{j'}^{*} +
\sum\limits_{n} V_{nj'}^{*}\beta_{j}\beta_{n}^{*} \right) \nonumber \\
& & \quad + \dfrac{i\Omega}{2}\sum\limits_{j}
\left( e^{i\textbf{k}_0\cdot\textbf{r}_j}\beta_j^{*}-c.c. \right) \nonumber \\
&=& \dfrac{\Gamma}{2} \sum\limits_{j}\sum\limits_{j'} \left( V_{jj'} + V_{jj'}^{*} \right) \beta_{j}\beta_{j'}^{*} \nonumber \\
& & \quad - \dfrac{i\Omega}{2}\sum\limits_{j}
\left( e^{-i\textbf{k}_0\cdot\textbf{r}_j}\beta_j-c.c. \right)
\label{I_2_final}
\end{eqnarray}

The first term is proportional to Eq.~(\ref{A_1}) after substituting
$V_{jj'} = e^{ik_{0}r_{jj'}}/(ik_{0}r_{jj'})$ and noting that it is symmetric to the exchange $j\leftrightarrows j'$. Furthermore, from Eq.~(\ref{I_1}), it is also equal to the total emitted power $P(t)$. Thus, we rewrite Eq.~(\ref{I_2_final}) as
\begin{equation}
P(t) \propto -\dfrac{d}{dt} \sum\limits_{j} \left| \beta_{j}(t) \right|^{2} + \dfrac{i\Omega}{2}\sum\limits_{j}
\left( e^{-i\textbf{k}_0\cdot\textbf{r}_j}\beta_j-c.c. \right)
\label{P_omega}
\end{equation}
for all $t$.

When the driving laser is switched off, $\Omega=0$ and Eq.~(\ref{P_omega}) yields to Eq.~(\ref{P_derivative}), valid during the decay.


\end{document}